\documentclass{article}
\usepackage{amssymb}
\usepackage{latexsym}
\usepackage{amsthm}
\usepackage{amsmath}
\begin{document}
\begin{titlepage}

    \thispagestyle{empty}

\begin{center}
        { \huge{\textbf{Adams-Iwasawa $\mathbf{\mathcal{N}=8}$ Black Holes}}}\let\thefootnote\relax\footnotetext{Contribution to the Proceedings of the `JW2011 Workshop on the Scientific and Human Legacy of Julius Wess', held August 27 - 28, 2011 in Donji Milanovac, Serbia}

 \vspace{20pt}

        {\large{\bf Sergio L. Cacciatori$^{1,4}$, Bianca L. Cerchiai$^{2,4}$, and \ Alessio Marrani$^3$}}

        \vspace{40pt}

        {$1$ \it Dipartimento di Scienze ed Alta Tecnologia,\\Universit\`a degli Studi dell'Insubria,
Via Valleggio 11, 22100 Como, Italy\\
\texttt{sergio.cacciatori@uninsubria.it}}

        \vspace{10pt}

        {$2$ \it Dipartimento di Matematica,\\
Universit\`a degli Studi di Milano,  Via Saldini 50, 20133 Milano,
Italy\\
\texttt{bianca.cerchiai@unimi.it}}
        \vspace{10pt}

        {$3$ \it Physics Department,Theory Unit, CERN, \\
        CH 1211, Geneva 23, Switzerland\\
        \texttt{alessio.marrani@cern.ch}}

\vspace{10pt}

        {$4$ \it INFN, Sezione di Milano\\
Via Celoria, 16, 20133 Milano,
Italy}

        \vspace{30pt}
\end{center}

\vspace{5pt}

\begin{abstract}
We study some of the properties of the geometry of the exceptional Lie group $E_{7(7)}$, which describes the U-duality of the $\mathcal{N}=8$, $d=4$ supergravity. In particular, based on a symplectic construction of the Lie algebra $\mathfrak{e}_{7\left( 7\right) }$ due to Adams, we compute the Iwasawa decomposition of the symmetric space $\mathcal{M}=\frac{E_{7\left( 7\right) }}{(SU\left(8\right)/\mathbb{Z}_2)}$, which gives the vector multiplets' scalar manifold of the corresponding supergravity theory.

The explicit expression of the Lie algebra is then used to analyze the origin of $\mathcal{M}$ as scalar configuration of the ``large" $\frac{1}{8}$-BPS extremal black hole attractors. In this framework it turns out that the $U(1)$ symmetry spanning such attractors is broken down to a discrete subgroup $\mathbb{Z}_4$, spoiling their dyonic nature near the origin
of the scalar manifold.

This is a consequence of the fact that the maximal manifest off-shell symmetry of the Iwasawa parametrization is determined by a completely non-compact Cartan subalgebra of the maximal subgroup $SL(8,\mathbb{R})$ of $E_{7\left( 7\right) }$, which breaks down the maximal possible covariance $SL(8,\mathbb{R})$ to a smaller $SL(7,\mathbb{R})$ subgroup.
These results are compared with the ones obtained in other known bases, such as the Sezgin-van Nieuwenhuizen and the Cremmer-Julia /de Wit-Nicolai frames.

\end{abstract}
\end{titlepage}
\section{Introduction}

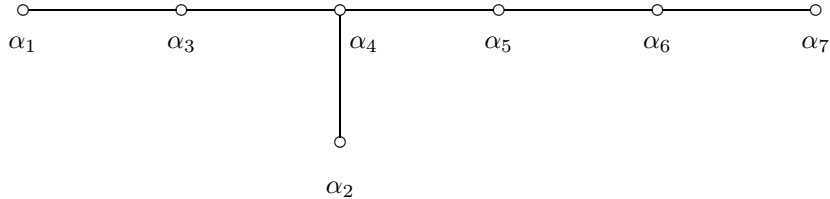
\begin{figure}[ph]
\begin{center}
\begin{picture}(100, 85)%
   \put(-90, 80){\circle{4}}%
   \put(-88, 80){\line(1, 0){56}}%
   \put(-30, 80){\circle{4}}%
   \put(-28, 80){\line(1, 0){56}}%
   \put(30, 80){\circle{4}}%
   \put(32, 80){\line(1, 0){56}}%
   \put(90, 80){\circle{4}}%
   \put(92, 80){\line(1, 0){56}}%
    \put(150, 80){\circle{4}}%
   \put(152, 80){\line(1, 0){56}}%
   \put(210, 80){\circle{4}}%
   \put(30, 78){\line(0, -1){46}}%
   \put(30, 30){\circle{4}}%
    \put(-90, 70){\makebox(0, 0)[t]{$\alpha_1$}}%
    \put(-30, 70){\makebox(0, 0)[t]{$\alpha_3$}}%
    \put(39, 70){\makebox(0, 0)[t]{$\alpha_4$}}%
    \put(90, 70){\makebox(0, 0)[t]{$\alpha_5$}}%
    \put(150, 70){\makebox(0, 0)[t]{$\alpha_6$}}%
   \put(210, 70){\makebox(0, 0)[t]{$\alpha_7$}}%
\put(30, 15){\makebox(0, 0)[t]{$\alpha_2$}}
\end{picture}
\end{center}
\vspace{-3.5ex}
\caption{Dynkin diagram of $\mathfrak{e}_7$}
\label{fig1}
\end{figure}

The exceptional Lie group $E_7$ plays an important role in supergravity~\cite{CJ,dWN} as well as in quantum information theory~\cite{Duff-QIT,4-qubits,Cerchiai-VG}. It is a simple exceptional Lie group of rank~7 and dimension~133. The complex algebra $\frak{e}_{7}$ is completely characterized by its Dynkin diagram as shown in Fig~\ref{fig1}.

The Lie algebra $\frak{e}_{7}$ admits four distinct real forms,
one for each possible choice of a maximal compact subgroup. In Table~\ref{tab1} we list them together with their maximal compact subgroup (mcs). In the case of the three non compact forms we also recall the supergravity theory for which they describe the electric magnetic duality according to the framework developed in Ref.~\cite{GZ}.
The second subscript in the name of the groups indicates the difference between the number of non compact (nc) and compact generators.

\begin{table}[ph]
\caption{Real forms of $E_7$}
\begin{tabular}{|c|c|c|c|c|c|}
\hline
&&&&& \\[-1ex]
Name  & Real form $G$ & \parbox{23ex}{Maximal compact\\ subgroup $K$=mcs($G$)} & nc & rank($\frac{G}{K}$) & Supergravity \\[2ex]  \hline
&&&&& \\[-1ex]
compact & $E_{7(-133)}$  & $E_{7(-133)}$  & 0 & 0 & \\[1ex]  \hline
&&&&& \\[-1ex]
\parbox{15ex}{minimally non compact} & $E_{7(-25)}$ & $\displaystyle{\frac{E_6 \times U(1)}{\mathbb{Z}_3}}$ & 54& 3 & \parbox{7ex}{$\mathcal{N}=2$\\ d=4}\\[3ex] \hline &&&&& \\[-1ex]
$E \, \,  V\! I$ & $E_{7(-5)}$ & $\displaystyle{\frac{Spin(12) \times USp(2)}{\mathbb{Z}_2}}$ & 64 & 4  & \parbox{7ex}{\mbox{$\mathcal{N}=4 \Leftrightarrow$} \mbox{$\mathcal{N}=12$}\\ d=3}\\[4ex]  \hline
&&&&& \\[-1ex]
split & $E_{7(7)}$ & $\displaystyle{\frac{SU(8)}{\mathbb{Z}_2}}$ & 70 & 7 & \parbox{7ex}{$\mathcal{N}=8$\\ d=4}\\[2ex] \hline
\end{tabular}
\label{tab1}
\end{table}

In particular, here we focus on the split (maximally non compact) form $E_{7(7)}$ relevant for $\mathcal{N}=8$ supergravity in 4 dimensions. In the past couple of years this theory has turned out to be increasingly interesting due to its remarkable convergence properties in the ultraviolet~\cite{UV}. Analyzing the geometry of the $E_{7(7)}$ symmetry can help to understand what is behind this.

First, in Sec.~\ref{sec2} following a symplectic construction developed in Ref.~\cite{adams}, we compute the Lie algebra $\mathfrak{e}_{7(7)}$ based on a $\mathfrak{sl}(8,\mathbb{R})$ subalgebra.
Then, in Sec.~\ref{sec3} we pick a completely non compact Cartan subalgebra of $\mathfrak{sl}(8,\mathbb{R})$ as a pivot to perform the Iwasawa decomposition of the coset manifold $\mathcal{M}=\frac{E_{7\left( 7\right) }}{(SU\left(8\right)/\mathbb{Z}_2)}$.
Finally, in Sec.~\ref{sec4} we recast the equations of $\mathcal{N}=8$ $d=4$ supergravity in the Adams-Iwasawa framework, obtaining explicit expressions for the central charge matrix $Z_{ij}$ and the effective black holes potential $V_{BH}$, which we apply to the analysis of the attractor mechanism. At the Lie algebra level we study the ``large" $\frac{1}{8}$-BPS attractor solutions at the origin of the scalar-manifold itself. Due to the breaking of a residual ``degeneracy symmetry'' $U\left( 1\right) $ down to a subgroup $\mathbb{Z}_{4}$ determined by the choice of the Cartan subalgebra, we find that the dyonic nature of such solutions is spoiled~\cite{e77}.

\section{\label{sec2} Adams symplectic construction of the Lie algebra $\mathbf{\mathfrak{e}_{7(7)}}$}

The Lie algebra $\frak{e}_{7(7)}$ can be constructed following Chapter 12 of Ref.~\cite{adams}.

Let $V$ be an $8$-dimensional real vector space, which we are going to identify with $\mathbb{R}^8$ later, and $V^{\ast }$ its dual. Let $\Lambda ^{i}V$ denote the $i$-th external power of $V$. Then $SL(V)$ can be defined as the group of automorphisms preserving the isomorphism $\Lambda ^{8}V\simeq \mathbb{R}$ and the Lie algebra $L=\mathfrak{sl}(V)$ of $SL(V)$ acts on the $56$-dimensional real vector space $W\equiv \Lambda ^{2}V\oplus \Lambda ^{2}V^{\ast }$ through:
\begin{equation}
L(W)=L(V)\wedge V\oplus L(V^{\ast })\wedge V^{\ast }+V\wedge L(V)\oplus V^{\ast }\wedge L(V^{\ast }),
\end{equation}
with $L(V^{\ast })$ the adjoint action. Moreover, for $i+j=8$ the isomorphism $\Lambda ^{i}V\simeq \Lambda ^{j}V^{\ast }$ induced by the wedge product~$\wedge$ yields the maps:
\begin{equation}
\Lambda ^{4} V \otimes \Lambda ^{2}V\overset{\wedge }{\longrightarrow }\Lambda ^{6}V\simeq \Lambda ^{2}V^{\ast };  \Lambda ^{4} V \otimes \Lambda ^{2}V^{\ast }\simeq \Lambda ^{4}V^{\ast
}\otimes \Lambda ^{2}V^{\ast }\overset{\wedge }{\longrightarrow }\Lambda
^{6}V^{\ast }\simeq \Lambda ^{2}V,
\label{maps}
\end{equation}
which allow us to define an action of $\Lambda ^{4} V$ on $W$.
Since $\Lambda ^{4} V$ has $dim_{\mathbb{R}}=\binom{8}{4}=70$, it follows that $A\equiv \mathfrak{sl}(V) \oplus \Lambda ^{4} V$ is a $133$-dimensional real vector space of operators acting on $W$. Then Theorem 12.1 of Ref.~\cite{adams} states that, up to isomorphisms, $A$ is a Lie algebra acting on $W$ in the same way as $\frak{e}_{7(7)}$ acts on its fundamental irrep. $\mathbf{56}$. And now we are going to implement this construction to explicitly compute it.

First, by identifying $V$ with $\mathbb{R}^{8}$, we can easily calculate a realization of $\mathfrak{sl}(8,\mathbb{R})$ in its irrep. $\mathbf{8}$, which we can subsequently use as a building block to obtain a matrix realization of the action of $\frak{sl}(V)$ on $\Lambda ^{2}V$ and afterwards on $W$, according to the above procedure.

Given a basis $\{e_{i}\}_{i=1,...,8}$ of $\mathbb{R}^{8}$, the algebra $\mathfrak{sl}(8,\mathbb{R})$ is generated by the $8\times 8-1=63$ traceless $8\times 8$ matrices $\{A_{kl},S_{kl},D_{\alpha }\}$ in $M(8,\mathbb{R})$, defined as follows:
\begin{eqnarray}
&&A_{kl}e_{i} \equiv \delta _{li}e_{k}-\delta _{ki}e_{l}=A_{[kl]}e_{i}; \quad
S_{kl}e_{i} \equiv \delta _{li}e_{k}+\delta _{ki}e_{l}=S_{(kl)}e_{i};
\mbox{with } 1\leq k<l\leq 8 \notag \\
&&D_{\alpha } \equiv \text{diag}\{D_{\alpha }^{1},\ldots ,D_{\alpha
}^{8}\};~~Tr\left( D_{\alpha }\right) =0; \mbox{with } 1 \leq \alpha \leq 7.
\label{8}
\end{eqnarray}
Here the $A_{kl}$'s and $S_{kl}$'s are $28$ antisymmetric and $28$ symmetric $8\times 8$ matrices, respectively, while the $D_{\alpha }$'s are $7$ diagonal traceless $8\times 8$ matrices. At this point some remarks are in order. Despite their traceless symmetry, the $S_{kl}$'s are only $28$ because of the index ordering $k<l$ and the remaining $8-1=7$
traceless diagonal degrees of freedom are implemented through the $D_{\alpha }$'s. In first line of (\ref{8}) the square brackets mean
antisymmetrization $\left( A_{[kl]}\equiv \frac{1}{2}(A_{kl}-A_{lk}) \right)$, whereas the round brackets indicate symmetrization $\left(S_{(kl)}\equiv \frac{1}{2}(S_{kl}+S_{lk})\right)$. Finally, the normalization is chosen such that:
\begin{equation}
Tr\left( A_{kl}A_{mn}\right) \equiv -2\delta _{kl\mid mn}; \quad Tr\left( S_{kl}S_{mn}\right) \equiv 2\delta _{kl\mid mn}; \quad Tr\left( D_{\alpha }D_{\beta }\right) \equiv 2\delta _{\alpha \beta }.
\end{equation}

Picking the basis $\left\{e_{ij}\right\} _{i<j}\equiv e_{i}\wedge e_{j}$ for $\Lambda ^{2}V$, the extension of the action of $\mathfrak{sl}(V)$ to $\Lambda ^{2}V$ is:
\begin{eqnarray}
&&A_{kl}(e_{ij})=\sum_{m,n}(U_{klim}^{A}D_{kljn}+D_{klim}U_{kljn}^{A})e_{mn};
\label{PPPA-1} \\
&&S_{kl}(e_{ij})=\sum_{m,n}(U_{klim}^{S}D_{kljn}+D_{klim}U_{kljn}^{S})e_{mn};
\label{PPPA-2} \\
&&D_{\alpha }(e_{ij})=(D_{\alpha }^{i}+D_{\alpha }^{j})e_{ij},
\label{PPPA-3}
\end{eqnarray}
and then choosing a dual basis $\{\varepsilon ^{ij}\}_{i<j}$ for $\Lambda ^{2}V^{\ast }$, we get:
\begin{eqnarray}
&&A_{kl}(\varepsilon
^{ij})=\sum_{m,n}(U_{klim}^{A}D_{kljn}+D_{klim}U_{kljn}^{A})\varepsilon
^{mn};  \label{1} \\
&&S_{kl}(\varepsilon
^{ij})=-\sum_{m,n}(U_{klim}^{S}D_{kljn}+D_{klim}U_{kljn}^{S})\varepsilon
^{mn};  \label{2} \\
&&D_{\alpha }(\varepsilon ^{ij})=-(D_{\alpha }^{i}+D_{\alpha
}^{j})\varepsilon ^{ij}.  \label{3}
\end{eqnarray}
Notice that in the sums we do not restrict $m<n$, rather we take into account that $e_{mn}=-e_{nm}$. Also, we are using the notation:
\begin{equation}
\begin{array}{l}
U_{klim}^{A} \equiv \delta _{km}\delta _{li}-\delta _{ki}\delta _{lm};\\[1ex]
U_{klim}^{S} \equiv \delta _{km}\delta _{li}+\delta _{ki}\delta _{lm};
\end{array}
\quad
D_{klim} \equiv \left\{
\begin{array}{l}
\delta _{im}~\text{for}~k\neq l\neq i; \\
0~\text{otherwise}.
\end{array}
\right.
\end{equation}
Equations (\ref{PPPA-1}), (\ref{1}) together,  and (\ref{PPPA-2}), (\ref{2}) together define the $56 \times 56$ matrices representing the action on $W$ of the operators $A_{kl}$ and $S_{kl}$ respectively. Analogously,  (\ref{PPPA-3}) and (\ref{3}) together define the $56 \times 56$ matrices $h_{D_\alpha}$ corresponding to the diagonal operators $D_\alpha$.

The remaining $70$ generators of $\frak{e}_{7(7)}$ spanning $\Lambda ^{4}V$ are obtained from the action of
$\underset{\left( i_{1}<i_{2}<i_{3}<i_{4}\right) }{\lambda
_{i_{1}i_{2}i_{3}i_{4}}}\equiv e_{i_{1}}\wedge e_{i_{2}}\wedge
e_{i_{3}}\wedge e_{i_{4}}$ on $W$ through the maps (\ref{maps}):
\begin{eqnarray}
&&(e_{i_{1}}\wedge e_{i_{2}}\wedge e_{i_{3}}\wedge e_{i_{4}})\otimes
(e_{j_{1}j_{2}})\mapsto \frac{1}{2}\epsilon
_{i_{1}i_{2}i_{3}i_{4}j_{1}j_{2}k_{1}k_{2}}\varepsilon ^{k_{1}k_{2}}; \\
&&(e_{i_{1}}\wedge e_{i_{2}}\wedge e_{i_{3}}\wedge e_{i_{4}})\otimes
(\varepsilon ^{j_{1}j_{2}})\mapsto \frac{1}{2}\delta
_{i_{1}i_{2}i_{3}i_{4}}^{j_{1}j_{2}k_{1}k_{2}}e_{k_{1}k_{2}},
\end{eqnarray}
where $\epsilon_{i_{1}i_{2}i_{3}i_{4}j_{1}j_{2}k_{1}k_{2}}$ is the standard  $8$-dimensional Levi-Civita tensor, and
$\delta _{i_{1}i_{2}i_{3}i_{4}}^{j_{1}j_{2}j_{3}j_{4}}\equiv \sum_{\sigma \in {\mathcal{P}}[1,2,3,4]}\epsilon _{\sigma }\delta _{i_{\sigma(1)}}^{j_{1}}\delta _{i_{\sigma (2)}}^{j_{2}}\delta _{i_{\sigma
(3)}}^{j_{3}}\delta _{i_{\sigma (4)}}^{j_{4}}$,
with ${\mathcal{P}}[1,2,3,4]$ the set of permutations of $[1,2,3,4]$
and $\epsilon _{\sigma }$ the parity of permutation $\sigma$.

Introducing a double-index notation for the matrices according to \linebreak \mbox{$(Me)^{ij}=\sum_{k<l}M^{ij\mid kl}e_{kl}$}, and switching to
block matrix form with respect to the decomposition $W\equiv \Lambda ^{2}V\oplus \Lambda ^{2}V^{\ast }$, the expression for $\lambda _{i_{1}i_{2}i_{3}i_{4}}$ becomes:
\begin{equation}
\lambda _{i_{1}i_{2}i_{3}i_{4}}= \left(
\begin{array}{cc}
0 & \epsilon _{i_{1}i_{2}i_{3}i_{4}ijkl} \\
\delta _{i_{1}i_{2}i_{3}i_{4}}^{ijkl} & 0
\end{array}
\right). \label{MI-1}
\end{equation}

In order to simplify the analysis of the covariance properties of the parametrization we need to carry out later in Sec.~\ref{sec3}, it is convenient to introduce the tetra-indices $I\equiv \left[i_{1}i_{2}i_{3}i_{4}\right]$, which are completely
antisymmetric and obey to the ordering rule $i_{1}<i_{2}<i_{3}<i_{4}$. This in turn uniquely determines the complementary tetra-index $\tilde{I}$, satisfying $\epsilon _{I\tilde{I}}\neq 0$ and $\left( \lambda _{I}\right) ^{T}=\epsilon _{I\tilde{I}}\lambda _{\tilde{I}}$.
Then we can perform the change of basis for $\Lambda ^{4}V$:
\begin{equation}
{}\mathcal{S}_{I}\equiv \frac{1}{\sqrt{2}}(\lambda _{I}+\epsilon _{I\tilde{I}%
}\lambda _{\tilde{I}}); \quad
\mathcal{A}{}_{I}\equiv \frac{1}{\sqrt{2}}(\lambda _{I}-\epsilon _{I\tilde{I}%
}\lambda _{\tilde{I}}),
\end{equation}
where the matrices $\mathcal{S}_{I}$'s are symmetric, while the $\mathcal{A}{}_{I}$'s are antisymmetric.
To avoid a double over-counting due to the fact that ${}\mathcal{S}_{I}={}\mathcal{S}_{\widetilde{I}}$ and $\mathcal{A}{}_{I}=-\mathcal{A}{}_{\widetilde{I}}$, we need to restrict the set of
tetra-indices. A consistent basis for $\Lambda ^{4}V$ is provided by the subset $\mathcal{I}_{8}\equiv \left[ ijk8\right] $, with $1\leq i<j<k\leq 7$, i.e. by the subset of  tetra-indices $\mathcal{I}$ with $i_4=8$. It has cardinality $\binom{7}{3}=\frac{1}{2}\binom{8}{4}=35$. Therefore, a good choice of a basis for $\Lambda ^{4}V$ reads:
$\{{}\mathcal{S}_{I},{}\mathcal{A}{}_{I}\}_{I\in {}\mathcal{I}_{8}}$.

Putting it all together, a realization of the algebra $\mathfrak{e}_{7(7)}$ in its irrep. $\mathbf{56}$ is given by the following $133$ orthogonal matrices:
\begin{equation}
\{A_{kl},{}\mathcal{A}_{I},h_{D_{\alpha }},S_{kl},\mathcal{S}_{I}\}, \mbox{ with }
1\leq k<l\leq 8; \,
1\leq \alpha \leq 8; \,
I\in \mathcal{I}_{8}.
\label{basis}
\end{equation}
The set of antisymmetric matrices $A_{\mu }\equiv \{A_{kl},%
\mathcal{A}_{I}\}$ is normalized as \mbox{$Tr\left( A_{\mu }A_{\nu }\right)=-2\delta _{\mu \nu }$}, and it has cardinality $28+35=63$ ($\mu =1,...,63$), so that $A_{\mu }$ generates the maximal compact (symmetric) subgroup $\frac{SU(8)}{\mathbb{Z}_2}$ of $E_{7\left( 7\right) }$ (see e.g. Ref.~\cite{Gilmore}). The set of the remaining $7+28+35=70$ symmetric matrices $S_{\Lambda }\equiv \{h_{D_{\alpha}},S_{kl},{}\mathcal{S}_{I}\}$ ($\Lambda =1,...,70$) is normalized as $%
Tr\left( S_{\Lambda }S_{M}\right) =2\delta _{\Lambda M}$, so that it spans
the non compact coset \mbox{$\mathcal{M}=\frac{E_{7\left( 7\right) }}{(SU\left(8\right)/\mathbb{Z}_2)}$}.

\section{\label{sec3} Iwasawa Decomposition of $\mathbf{\mathcal{M}=\frac{E_{7\left( 7\right) }}{(SU\left(8\right)/\mathbb{Z}_2)}}$}

The symmetric space $\mathcal{M}=\frac{E_{7\left( 7\right) }}{(SU\left(8\right)/\mathbb{Z}_2)}$ has rank $7$ (see Table~\ref{tab1}), which means that it is possible to find a non compact Cartan subalgebra completely outside of $\frak{su(8)}$. This is a consequence of the fact that $E_{7(7)}$ is the split form of $E_7$. We can pick such a Cartan subalgebra $\mathcal{C}$ as the algebra generated by the $7$ diagonal matrices $h_{D_{\alpha }}$ defined through (\ref{PPPA-3}) and (\ref{3}):
\begin{equation}
\mathcal{C}\equiv \left\langle h_{D_{\alpha }}\equiv \left(
\begin{array}{cc}
(D_{\alpha }^{i}+D_{\alpha }^{j})\delta _{ij}^{mn} & 0 \\
0 & -(D_{\alpha }^{i}+D_{\alpha }^{j})\delta _{mn}^{ij}
\end{array}
\right)\right\rangle _{\mathbb{R}}\subsetneq \frak{e}_{7(7)}.
\label{cartan}
\end{equation}

As the next step, we need to choose a complete set of positive roots of $\frak{e}_{7(7)}$ with respect to $\mathcal{C}$. Such a choice is provided by the set $J^{+}\cup \mathcal{J}$ $^{+}$, where
\begin{eqnarray}
 J^{+} &\equiv &\{J_{kl}^{+}\equiv \frac{1}{\sqrt{2}}(S_{kl}+A_{kl})=\sqrt{2}\left(
\begin{array}{cc}
\delta _{li}\delta _{kj}^{mn}-\delta _{lj}\delta _{ki}^{mn} & 0 \\
0 & \delta _{kn}\delta _{lm}^{ij}-\delta _{km}\delta _{ln}^{ij}
\end{array}
\right)\,|\,k<l\}; \notag \\
\mathcal{J}^{+}&\equiv &\{\mathcal{J}_{I}^{+} \equiv \lambda _{I} = \left(
\begin{array}{cc}
0 & \epsilon _{i_{1}i_{2}i_{3}i_{4}ijmn} \\
\delta _{i_{1}i_{2}i_{3}i_{4}}^{ijmn} & 0
\end{array}
\right) \in \Lambda ^{4}V {}:I\in \mathcal{I}%
_{8}\}.
\label{JJ}
\end{eqnarray}
Here we have applied the definitions (\ref{PPPA-2}) and (\ref{2}) for $S_{kl}$, (\ref{PPPA-1}) and (\ref{1}) for $A_{kl}$, and (\ref{MI-1}) for $\lambda_I$ respectively. Notice that $J_{kl}^{+}$ and $\mathcal{J}_{I}^{+}$ are nilpotent, as expected:
\begin{equation}
(J_{kl}^{+})^{2}=0; \quad (\mathcal{J}_{I}^{+})^{2}=0.
\end{equation}
Finally, an Iwasawa parametrization for the representative of the irreducible, Riemannian, globally symmetric coset space $\mathcal{M}=\frac{E_{7\left( 7\right) }}{(SU\left(8\right)/\mathbb{Z}_2)}$ can be constructed as~\cite{e77}:
\begin{equation}
\mathbf{C}(x^{\alpha },x^{ij},x^{I})=\exp \left( \sum_{\alpha
=1}^{7}x^{\alpha }h_{D_{\alpha }}\right) \prod_{i<j}\exp \left(
x^{ij}J_{ij}^{+}\right) \prod_{I\in \mathcal{I}_{8}}\exp \left( x^{I}%
\mathcal{J}_{I}^{+}\right) .  \label{coset}
\end{equation}

Before we are able to proceed with the application to $\mathcal{N}=8$ $d=4$ supergravity, a comment on the manifest covariance properties of the $70$ real scalars $\phi \equiv \left\{ x^{\alpha},x^{kl},x^{I}\right\} $ in our parametrization is needed. The scalars have different types of indices, and thus different covariance properties, namely:
\begin{itemize}
\item  $I\in \mathcal{I}_{8}$ is in the rank-$3$ antisymmetric irreducible
representation $\mathbf{35}$ of $SL\left( 7,\mathbb{R}\right) $;

\item  $kl=\left[ kl\right] $ is in the rank-$2$ antisymmetric
(contra-gradient) $\mathbf{28}^{\prime }$ irrep. of $SL\left( 8,\mathbb{R}
\right) $;

\item  $\alpha $ is in the fundamental irrep. $\mathbf{7}$ of $SL\left( 7,%
\mathbb{R}\right) $.
\end{itemize}

Thus, the maximal common covariance of this framework is $SL\left( 7,\mathbb{R}\right) $, breaking the maximal possible off-shell covariance $SL\left(8,\mathbb{R}\right) $ of $\mathcal{N}=8$ $d=4$ supergravity:
\begin{equation}
SL\left( 8,\mathbb{R}\right) \supset SL\left( 7,\mathbb{R}\right) \times
SO\left( 1,1\right) \quad \mbox{ according to } \quad
\mathbf{28}^{\prime }= \mathbf{21}_{1}^{\prime }+\mathbf{7}%
_{-3}^{\prime },
\label{br-1}
\end{equation}
where $\mathbf{21}$ is the rank-$2$ antisymmetric (contra-gradient) irrep.
of $SL\left( 7,\mathbb{R}\right) $, and the subscripts denote the weights
with respect to $SO\left( 1,1\right) $.

\section{\label{sec4} Iwasawa $\mathbf{\mathcal{N}=8}$ Supergravity}

In this Section we are reformulating  $\mathcal{N}=8$, $d=4$ ungauged supergravity theory in the symplectic Adams-Iwasawa frame,
computing the central charge matrix $Z_{ij}$ and the effective black holes potential $V_{BH}$ in terms of the symplectic electric and magnetic sections. To this aim in order to compare our expressions e.g. with Ref.~\cite{CFGM1}, we need to recast the expression (\ref{coset}) for $\mathcal{M}$ in the block matrix form:
\begin{eqnarray}
\mathbf{C}\left( x^{\alpha },x^{ij},x^{I}\right)
&\equiv &\frac{1}{\sqrt{2}}\left(
\begin{array}{ccc}
\left( W_{1}\right) _{ij}^{~~mn} & ~ & \left( V_{1}\right) _{ij\mid mn} \\
~ & ~ & ~ \\
\left( V_{2}\right) ^{ij\mid mn} & ~ & \left( W_{2}\right) _{~~mn}^{ij}
\end{array}
\right),  \label{sections}
\end{eqnarray}
with $i,j=1,...,8$ in the fundamental irrepr. $\mathbf{8}$ of $SL\left(
8,\mathbb{R}\right) $, and all the indices antisymmetrized, ($ij=\left[ ij\right] $, $mn=\left[ mn\right]$ throughout). Notice that from the construction of the matrices $h_{D_{\alpha}}$ (\ref{cartan}) and $J_{kl}^{+}$ and $\mathcal{J}_{I}^{+}$ (\ref{JJ}) in Sec.~\ref{sec2}, this block matrix form (\ref{sections}) of the coset representative corresponds  to the branching:
\begin{equation}
E_{7\left( 7\right) }\supsetneq _{symm}^{\max }SL\left( 8,\mathbb{R}\right) \quad \mbox{ according to } \quad
\mathbf{56}= \mathbf{28}+\mathbf{28}^{\prime }.
\end{equation}
Thus, as observed in Sec.~\ref{sec3}, the maximal possible off-shell symmetry of the theory is $SL\left(8,\mathbb{R}\right)$, which has maximal compact subgroup: $SO(8)=mcs(SL\left(8,\mathbb{R}\right))=SU(8)\cap SL\left( 8,\mathbb{R}\right)$.
A complex manifestly $SU\left( 8\right) $-covariant $\frac{E_{7\left(
7\right) }}{(SU\left( 8\right)/\mathbb{Z}_2)}$-coset representative:
\begin{eqnarray}
\mathcal{V} &\equiv &\left(
\begin{array}{ccc}
u_{ij}^{~~mn} & ~ & v_{ij\mid mn} \\
~ & ~ & ~ \\
v^{ij\mid mn} & ~ & u_{~~mn}^{ij}
\end{array}
\right) \equiv R\,\,\mathbf{C}\left( x^{\alpha },x^{ij},x^{I}\right)
R^{-1}=  \notag \\
&& \label{V-call} \\
&=&\frac{1}{2\sqrt{2}}\left(
\begin{array}{ccc}
\left[ W_{1}+W_{2}+i\left( V_{2}-V_{1}\right) \right] _{ij}^{~~mn} & ~ &
\left[ V_{1}+V_{2}-i\left( W_{1}-W_{2}\right) \right] _{ij\mid mn} \\
~ & ~ & ~ \\
\left[ V_{1}+V_{2}+i\left( W_{1}-W_{2}\right) \right] ^{ij\mid mn} & ~ &
\left[ W_{1}+W_{2}+i\left( V_{1}-V_{2}\right) \right] _{~~mn}^{ij}
\end{array}
\right)  \notag
\end{eqnarray}
can be obtained by means of a Cayley rotation implemented by the unitary matrix $
R\equiv \frac{1}{\sqrt{2}}\left(
\begin{array}{ccc}
\mathbf{I} & ~ & i\mathbf{I} \\
~ & ~ & ~ \\
i\mathbf{I} & ~ & \mathbf{I}
\end{array}
\right).$
This is similar to what happens for the de Wit-Nicolai's basis~\cite{dWN}.
At this point, by applying the same procedure as in Eq. (3.2) of Ref.~\cite{CFGM1} and by relying on the symplectic formalism for extended supergravities recently reviewed e.g. in Ref.~\cite{ADFT-rev-1}, the electric and magnetic symplectic sections $\mathbf{f}$ and $\mathbf{h}$ of $\mathcal{N}=8$, $d=4$ supergravity are defined respectively as:
\begin{eqnarray}
\mathbf{f}_{ij}^{~~mn} &\equiv &\frac{1}{\sqrt{2}}\left( u+v\right)
_{ij}^{~~mn}=\frac{1}{4}\left[ \left( W_{1}+W_{2}+V_{1}+V_{2}\right)
+i\left( -W_{1}+W_{2}-V_{1}+V_{2}\right) \right] _{ij}^{~~mn};  \notag \\
\mathbf{h}_{ij\mid mn} &\equiv &-\frac{i}{\sqrt{2}}\left( u-v\right)
_{ij\mid mn}=\frac{1}{4}\left[ \left( W_{1}-W_{2}-V_{1}+V_{2}\right)
+i\left( -W_{1}-W_{2}+V_{1}+V_{2}\right) \right] _{ij\mid mn},  \notag
\end{eqnarray}
which means that, in turn, the $8\times 8$ antisymmetric central charge matrix $Z_{ij}$ (see e.g. Eq. (3.14) of Ref.~\cite{CFGM1} or Ref.~\cite{ADFT-rev-1}) can be constructed as:
\begin{eqnarray}
&&Z_{ij}\left( x^{\alpha },x^{kl},x^{I};q_{mn},p^{mn}\right) \equiv \mathbf{f}_{ij}^{~~mn}q_{mn}-\mathbf{h}_{ij\mid mn}p^{mn}=
\label{centralcharge} \\
&&=\frac{1}{4}\left[
\begin{array}{l}
\left( W_{1}+W_{2}+V_{1}+V_{2}\right) + \\
+i\left( -W_{1}+W_{2}-V_{1}+V_{2}\right)
\end{array}
\right] _{ij}^{~~mn} \! \! \! \! q_{mn}-\frac{1}{4}\left[
\begin{array}{l}
\left( W_{1}-W_{2}-V_{1}+V_{2}\right) +\\
+i\left( -W_{1}-W_{2}+V_{1}+V_{2}\right)
\end{array}
\right] _{ij\mid mn} \! \! \! \! \! p^{mn}.
\notag
\end{eqnarray}
Following Ref.~\cite{ADFT-rev-1} (see also Eq. (3.17) of Ref.~\cite{CFGM1}), now we can plug in the expression (\ref{centralcharge}) for $Z$ to calculate the positive definite effective black hole potential $V_{BH}$ \cite{FGK}:
\begin{eqnarray}
V_{BH} &\equiv &\frac{1}{2}Tr\left( ZZ^{\dag }\right) =\frac{1}{2}Z_{ij}%
\overline{Z}^{ij}=  \label{bhp} \\
&=&\frac{1}{2^{5}}\left[ \left( W_{1}+W_{2}+V_{1}+V_{2}\right) +i\left(
-W_{1}+W_{2}-V_{1}+V_{2}\right) \right] _{ij}^{~~mn}\cdot  \notag \\
&&\cdot \left[ \left( W_{1}+W_{2}+V_{1}+V_{2}\right) -i\left(
-W_{1}+W_{2}-V_{1}+V_{2}\right) \right] ^{~~ij\mid rs}q_{mn}q_{rs}+  \notag
\end{eqnarray}
\begin{eqnarray}
&&-\frac{1}{2^{5}}\left[ \left( W_{1}+W_{2}+V_{1}+V_{2}\right) +i\left(
-W_{1}+W_{2}-V_{1}+V_{2}\right) \right] _{ij}^{~~mn}\cdot  \notag \\
&&\cdot \left[ \left( W_{1}-W_{2}-V_{1}+V_{2}\right) -i\left(
-W_{1}-W_{2}+V_{1}+V_{2}\right)\right] _{~~rs}^{ij}q_{mn}p^{rs}+  \notag \\
&&  \notag \\
&&-\frac{1}{2^{5}}\left[ \left( W_{1}-W_{2}-V_{1}+V_{2}\right) +i\left(
-W_{1}-W_{2}+V_{1}+V_{2}\right) \right] _{ij\mid rs}\cdot  \notag \\
&&\cdot \left[ \left( W_{1}+W_{2}+V_{1}+V_{2}\right) -i\left(
-W_{1}+W_{2}-V_{1}+V_{2}\right) \right] ^{~~ijmn}q_{mn}p^{rs}+  \notag \\
&&  \notag \\
&&+\frac{1}{2^{5}}\left[ \left( W_{1}-W_{2}-V_{1}+V_{2}\right) +i\left(
-W_{1}-W_{2}+V_{1}+V_{2}\right) +\right] _{ij\mid mn}\cdot  \notag \\
&&\cdot \left[ \left( W_{1}-W_{2}-V_{1}+V_{2}\right) -i\left(
-W_{1}-W_{2}+V_{1}+V_{2}\right)\right] _{~~rs}^{ij}p^{mn}p^{rs}.  \notag
\end{eqnarray}

We want to study the properties of the ``large" $\frac{1}{8}$-BPS attractors in $\mathcal{N}=8$, $d=4$ supergravity, in particular around the origin of the scalar manifold $\mathcal{M}$:
\begin{equation}
x^{\alpha }=0;
x^{ij}=0;
x^{I}=0.
\label{origin}
\end{equation}
 It is known that the ``large" $\frac{1}{8}$-BPS orbit is given by \mbox{$\mathcal{O}_{\frac{1}{8}-BPS,\text{large}}=\frac{E_{7\left(-7\right) }}{E_{6\left( 2\right) }}$}~\cite{FG,LPS-1}
and that it has a moduli space \cite{ADF-U-duality-d=4,Ferrara-Marrani-2,CFGM1,ICL-1}
 $\mathfrak{M}_{\frac{1}{8}-BPS,\text{large}} =\frac{E_{6\left( 2\right) }}{SU\left( 6\right) \times SU\left( 2\right) }$.

The Attractor Eqs. are nothing but criticality conditions for the
effective black hole potential $V_{BH}$ \cite{FGK}:
\begin{equation}
\partial _{\phi }V_{BH}=0:\left. V_{BH}\right| _{\partial _{\phi
}V_{BH}=0}\neq 0.  \label{AEs}
\end{equation}
{From} Refs.~\cite{FG,FK-N=8,CFGM1} the origin (\ref{origin})
of the scalar manifold $\mathcal{M}$, as a $\frac{1}{8}$-BPS attractor solution, is supported by the skew-diagonal charge configuration:
\begin{equation}
Z_{ij,\frac{1}{8}-BPS,\text{large}}\equiv \frac{1}{2}
\left( q_{ij}+ip^{ij}\right)=e^{i\varphi /4}\text{diag}\left( r,0,0,0\right) \otimes \epsilon ,
\label{cc8}
\end{equation}
with $r\in \mathbb{R}_{0}^{+}$, $\varphi \in \left[ 0,8\pi \right) $ and $\epsilon$ the $2\times 2$ symplectic metric.
The expression (\ref{cc8}) exhibits a maximal compact symmetry
$SU\left( 6\right) \times SU\left( 2\right) =mcs\left( E_{6\left( 2\right)
}\right)$, corresponding to the following chain of maximal symmetric group embeddings~\cite{Gilmore}:
\begin{equation}
E_{7\left( 7\right) }\overset{mcs}{\supsetneq }SU\left( 8\right) \supsetneq
SU\left( 6\right) \times SU\left( 2\right) \times U\left( 1\right)_{\mathcal{A}} .
\label{emb-1a}
\end{equation}
It is important to notice that while $r$ is invariant with respect to $SU\left( 6\right) \times SU\left( 2\right) \times U(1)_{\mathcal{A}}$, $\varphi$ is not invariant under this $U(1)_{\mathcal{A}}$, which can thus be seen as a residual symmetry parametrizing the dyonic nature of $\frac{1}{8}$-BPS attractor solutions at the coset origin. We are denoting this $U(1)_{\mathcal{A}}$ subgroup with the subscript $\mathcal{A}$ to distinguish it from another $U(1)_{\mathcal{E}}$ subgroup  appearing in (\ref{emb2}).

Now let's proceed to the solution of the Attractor Eqs. (\ref{AEs}) around the origin (\ref{origin}) of $\mathcal{M}$ within the Adams-Iwasawa approach. It is enough to work at the Lie algebra level, i.e. to compute the terms of $V_{BH}$ (\ref{bhp}) which are linear in the scalar fields and to plug them into the Attractor Eqs. (\ref{AEs}). This yields the following system:
\begin{equation}
\left\{
\begin{array}{l}
(h_{D_{\alpha }})_{\ \ rs}^{mn}q_{mn}p^{rs}=0; \\[0.5ex]
(J_{kl}^{+}+J_{kl}^{+T})_{\ \ rs}^{mn}q_{mn}p^{rs}=0; \\[0.5ex]
(\epsilon +\delta )_{I}^{\ mnrs}q_{mn}q_{rs}-(\epsilon +\delta
)_{Imnrs}p^{mn}p^{rs}=0.
\end{array}
\right.  \label{Iwa-AEs}
\end{equation}
While (\ref{Iwa-AEs}) doesn't constrain $r$, it imposes the following condition for $\varphi$:
\begin{equation}
(D_{\alpha }^{1}+D_{\alpha }^{2})\sin \frac{\varphi }{2}=0,~~\forall \alpha
=1,...,7.  \label{sol-cond}
\end{equation}
But as $D_{\alpha }^{1}+D_{\alpha }^{2}$ cannot vanish for all $\alpha $
's, because the $D_{\alpha }$'s (\ref{8}) form a basis for the $8\times 8$ diagonal traceless matrices with real entries, this means that we are restricted to two types of solutions: the purely electric one with
$r=q$, $\varphi =4k\pi ,~k\in \mathbb{Z}$, and the purely magnetic one
with $r=p$, $\varphi =2\left( 2k+1\right) \pi ,~k\in \mathbb{Z}$.
In other words the $U(1)_{\mathcal{A}}$ symmetry has been broken down to a finite discrete $\mathbb{Z}_4$ subgroup.

This can be understood with the following reasoning.
As we have observed in Sec.~\ref{sec3}, due to the choice of a Cartan subalgebra the Adams-Iwasawa construction of the $\frac{E_{7\left( 7\right) }}{(SU\left(8\right)/\mathbb{Z}_2}$-coset representative $\mathbf{C}$ (\ref{coset}) explicitly breaks the maximal covariance from $SL(8,\mathbb{R})$ down to $SL(7,\mathbb{R})$  according to the branching (\ref{br-1}), which through the Cayley transformation turns into:
\begin{equation}
SU(8) \overset{mcs}{\supsetneq} SU(7) \times U(1)_{\mathcal{E}} \overset{mcs}{\supsetneq} SU(6)\times U(1)_{\mathcal{B}}\times U(1)_{\mathcal{E}}.
\label{emb2}
\end{equation}
But since the two subgroups $U(1)_{\mathcal{E}}$ in (\ref{emb2}) and $U(1)_{\mathcal{A}}$ in (\ref{emb-1a}) do not coincide, we are left with a residual discrete symmetry~\cite{e77}:
\begin{equation}
U(1)_{\mathcal{A}}\cap U(1)_{\mathcal{E}}=\mathbb{Z}_4 .
\end{equation}

Finally, let's compare the features of the Adams-Iwasawa approach to other known frameworks.
\begin{itemize}
\item  The Sezgin-van Nieuwenhuizen \cite{SVN,CFGM1} construction has
$USp(8)\subsetneq _{symm}^{\max }SU(8)$ as maximal manifest symmetry, which coincides with the maximal compact subgroup of the $\mathcal{N}=8$, $d=5$ $U$-duality group $E_{6\left( 6\right) }$: $USp(8)=mcs\left( E_{6\left(6\right) }\right) $. By recalling the explicit form of ``large'' non-BPS charge orbit in $\mathcal{N}=8$, $d=4$ supergravity \cite{FG,LPS-1}: $\mathcal{O}_{nBPS}=\frac{E_{7\left( 7\right) }}{E_{6\left( 6\right) }}$, it is clear that this basis provides the natural context for the investigation of ``large'' dyonic non-BPS $d=4$ extremal black holes.

\item  The Cremmer-Julia or de Wit-Nicolai \cite{CJ,dWN,CFGM1}
parametrization has a manifest covariance $SO(8)=mcs\left( SL\left( 8,\mathbb{R}\right) \right) \subsetneq _{symm}^{\max }E_{7(7)}$, providing a natural context in which $\frac{1}{8}-$BPS ``large'' extremal $d=4$ black holes can be treated~\cite{CFGM1}.

\item  The Adams-Iwasawa basis spoils the dyonic nature of the black hole attractors, providing a way to single out its purely electric or its purely magnetic component.
\end{itemize}

\section*{Acknowledgments}

The work of BLC has been supported in part by the European Commission under the FP7-PEOPLE-IRG-2008 Grant n PIRG04GA-2008-239412 \textit{``String Theory and Noncommutative Geometry''} (\textit{STRING})
and in part by a grant funded by the company LeSaffre Italia.

\end{document}